\documentclass[conference]{IEEEtran}
\usepackage[utf8]{inputenc}
\usepackage{dsfont}

\usepackage{config}
\usepackage{empheq}
\usepackage[noadjust]{cite}
\usepackage{booktabs}

\definecolor{myblue}{rgb}{.8, .8, 1}

\ifCLASSINFOpdf
  \usepackage[pdftex]{graphicx}
  \usepackage{epstopdf}
  \graphicspath{{./figures/}}
  \DeclareGraphicsExtensions{.eps}
\else
\fi

\author{\IEEEauthorblockN{Abdoulaye Tall\IEEEauthorrefmark{1}, Zwi Altman \IEEEauthorrefmark{1} and Eitan Altman\IEEEauthorrefmark{2}} \\ \IEEEauthorblockA{\IEEEauthorrefmark{1}Orange Labs
38/40 rue du General Leclerc,92794 Issy-les-Moulineaux \\Email: \{abdoulaye.tall,zwi.altman\}@orange.com}\\ \IEEEauthorblockA{\IEEEauthorrefmark{2}INRIA Sophia Antipolis, 06902 Sophia Antipolis, France, Email:eitan.altman@sophia.inria.fr}
}

\title{Virtual sectorization: design and self-optimization}
\begin{document}
\maketitle

\begin{abstract}

\ac{ViSn} aims at covering a confined area such as a traffic hot-spot using a narrow beam. The beam is generated by a remote antenna array located at- or close to the \ac{BS}. This paper develops the \ac{ViSn} model and provides the guidelines for designing the \ac{ViS} antenna. In order to mitigate interference between the \ac{ViS} and the traditional macro sector covering the rest of the area, a \ac{DSA} algorithm that self-optimizes the frequency bandwidth split between the macro cell and the \ac{ViS} is also proposed. The \ac{SON} algorithm is constructed to maximize the proportional fair utility of all the users throughputs. Numerical simulations show the interest in deploying \ac{ViSn}, and the significant capacity gain brought about by the self-optimized bandwidth sharing with respect to a full reuse of the bandwidth by the \ac{ViS}.


\end{abstract}

\section{Introduction}

	The increase of traffic demand has motivated the development of different solutions for increasing network capacity. \ac{AAS}, and in particular, \ac{VeSn}, has been one such solution \cite{yilmaz_selfoptimizationcoverage_2010}. The \ac{VeSn} consists of two vertically separated beams supporting two distinct sectors, denoted as inner and outer cells, transmitted by a single antenna. The inner cell is closed to the \ac{BS} and typically covers a small portion of the cell surface of the order of 20 percent or less. \ac{VeSn} is of interest when significant traffic is located at the inner cell coverage area. Different resource allocation strategies can be used such as full reuse of the frequency bandwidth by each of the sectors. One can further improve the performance of the system by intelligently activating \ac{VeSn} when traffic is present in the inner cell \cite{trichias_performanceevaluationson_2014}. Conversely, one can dynamically allocate frequency bandwidth in order to reduce interference which in turn maximizes the cell capacity \cite{tall_selfoptimizingstrategies_2015}. Such bandwidth allocation can be viewed as one possible dynamic implementation of the \ac{eICIC} as defined in the standard \cite{3gpp_evolveduniversalterrestrial_2013} in the frequency domain.

\begin{figure}[!ht]
\centering
\includegraphics[width=2.8in]{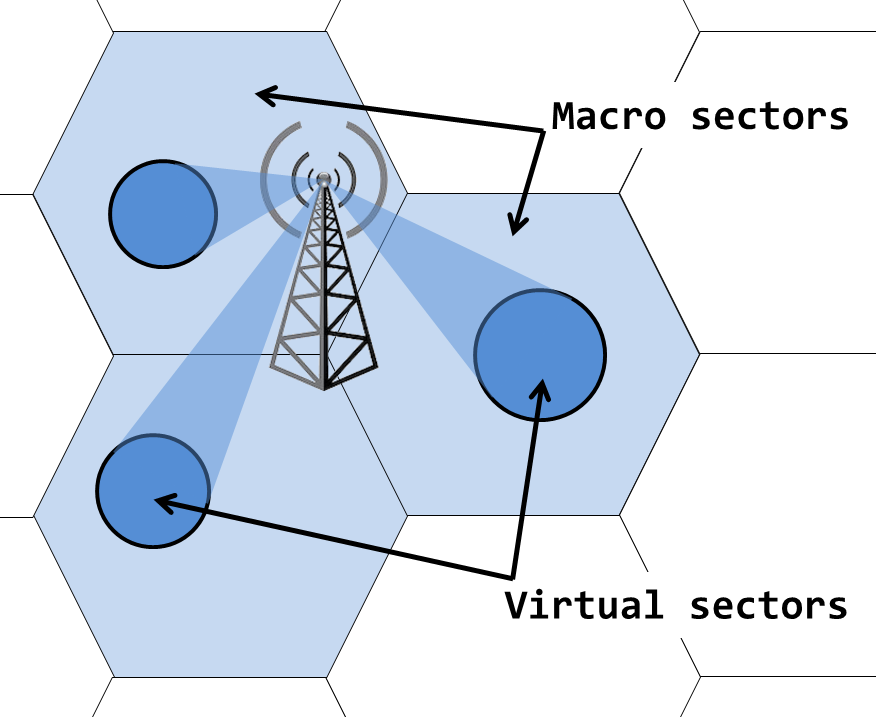}
\caption{Network layout with \ac{ViSn} enabled}
\label{fig:net_lay}
\end{figure}

	When the cell covers hot-spots which are located away from the \ac{BS}, \ac{VeSn} provides no advantages, and in this case, one can deploy small cells at the hot-spot area. The effectiveness of small cells grows when the hot-spot is located close to the cell edge. The deployment of backhaul can increase the overall cost of the small cell technology, particularly when optical backhaul is chosen. An alternative solution for small cell deployment is the use of large antenna array for generating narrow beams for covering the hot-spot's area in the cell as in Figure \ref{fig:net_lay}. A cell covered by a remote beam from an antenna located at- or near to the macro \ac{BS} is denoted as a \ac{ViS}.

	The purpose of this paper is to study different aspects of \ac{ViS} design and evaluate its performance. The study includes the antenna modeling and the resource allocation scheme which self-optimizes the cell capacity and performance. Two resource allocation schemes are possible. The first one consists in activating the \ac{ViS} with a full reuse of the same bandwidth as the macro cell, this scheme is denoted hereafter as \textit{bandwidth reuse one}. In this case, the macro cell and the \ac{ViS} share the total transmit power available at the \ac{BS}. The second scheme consists in sharing the total available bandwidth between the macro cell and the \ac{ViS}. We denote this scheme by \textit{bandwidth sharing} and we adopt the self-optimizing frequency splitting introduced in \cite{tall_selfoptimizingstrategies_2015} for the sharing proportions between the macro and the virtual cells.

	The paper is organized as follows. Section \ref{sec:antenna_design} describes the \ac{ViS} antenna model and guidelines for its design. Section \ref{sec:son_algo} presents the \ac{SON} algorithm for the resource allocation between the macro cell and the \ac{ViS} in its coverage area. The performance results of \ac{ViSn} are presented in Section \ref{sec:perf_results} followed by Section \ref{sec:conclusion} which concludes the paper.

\section{\ac{ViS} Antenna Design} \label{sec:antenna_design}

\begin{figure}[!ht]
\centering
\includegraphics[width=2.8in]{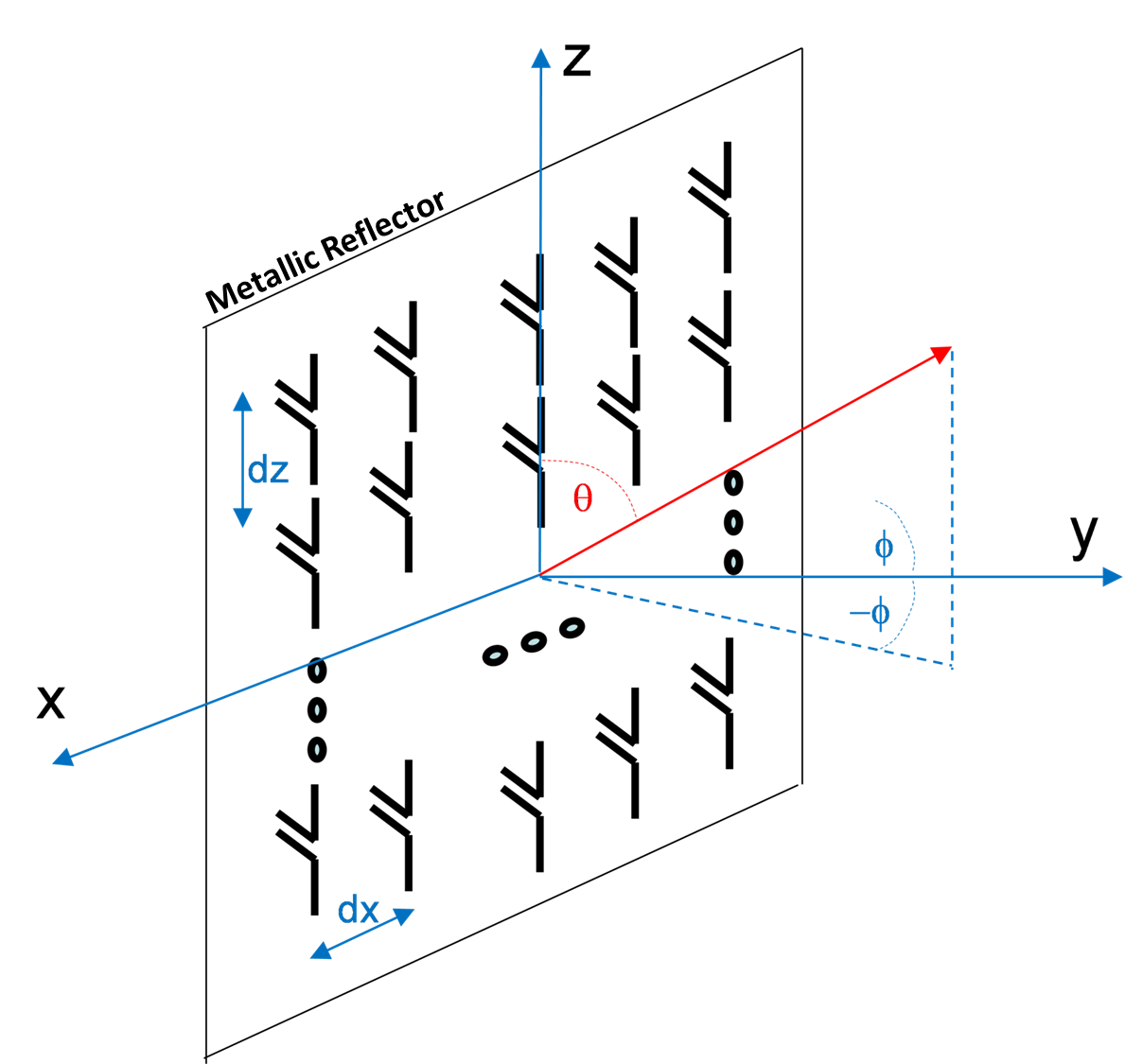}
\caption{\ac{ViS} antenna array}
\label{fig:ant_array}
\end{figure}

	This section provides the main guideline for the \ac{ViS} antenna design. The \ac{ViS} antenna comprises a two dimensional array with $N_x \times N_z$ elementary vertical dipoles in front of a metallic planar rectangular reflector. Other radiating elements can be chosen as well. It is noted that if another radiating element is chosen, its gain function should be modified while the rest of the model remains unchanged. $N_x$ and $N_z$  elements in each row and column respectively are equally spaced with distances $d_x$ and $d_z$  in the $x$ and $z$ directions respectively (Figure \ref{fig:ant_array}).

The antenna array creates a beam which covers the \ac{ViS} area. The beam direction is defined by the electrical tilt angles $\theta_e$  and $\phi_e$ in the spherical coordinates $\theta$ and $\phi$. The antenna gain is written as
	\begin{equation} \label{eq:ant_gain}
		G(\theta,\phi,\theta_e,\phi_e )=G_0 f(\theta,\phi,\theta_e,\phi_e )
	\end{equation}
	where $f$ is a normalized gain function and $G_0$ is the maximum gain. The excitation of the radiating dipoles is assumed to be separable in the $x$ and $z$ directions. Hence the function $f$ has the following form:
	\begin{equation} \label{eq:norm_gain_func}
	\begin{split}
	f(\theta,\phi,\theta_e,\phi_e ) =
	| \text{AF}_x^2 (\theta,\phi,\theta_e,\phi_e ) \cdot \text{AF}_y^2 (\theta,\phi) \\
	 \cdot \text{AF}_z^2 (\theta,\theta_e )| \cdot G_d (\theta)
	\end{split}
	\end{equation}
	where $\text{AF}_x$ and $\text{AF}_z$ are the array factors in the $x$ and $z$ directions respectively.

	The linear array is chosen with Gaussian tapering. The tapering provides larger weight to elements close to the center of the array, and consists of one lever for reducing the side lobe level. The term $\text{AF}_y (\theta,\phi)$ accounts for the impact of the metallic reflector. For sake of simplicity, we assume here an infinite perfect electric conductor at distance $\lambda/4$ from the dipoles. Hence $\text{AF}_y (\theta,\phi)$ can be written as
\begin{equation} \label{eq:array_factor}
	\text{AF}_y (\theta,\phi) = \sin(\frac{\pi}{2} \sin(\theta) \cos(\phi))
\end{equation}

	The term $G_0$ is obtained from power conservation equation:
	\begin{equation} \label{eq:max_gain}
		G_0 = \frac{4\pi}{\int_{-\pi/2}^{\pi/2} \int_0^\pi f(\theta,\phi) sin(\theta) d \theta d\phi}
	\end{equation}

\begin{figure}[!ht]
\centering
\includegraphics[width=2.8in]{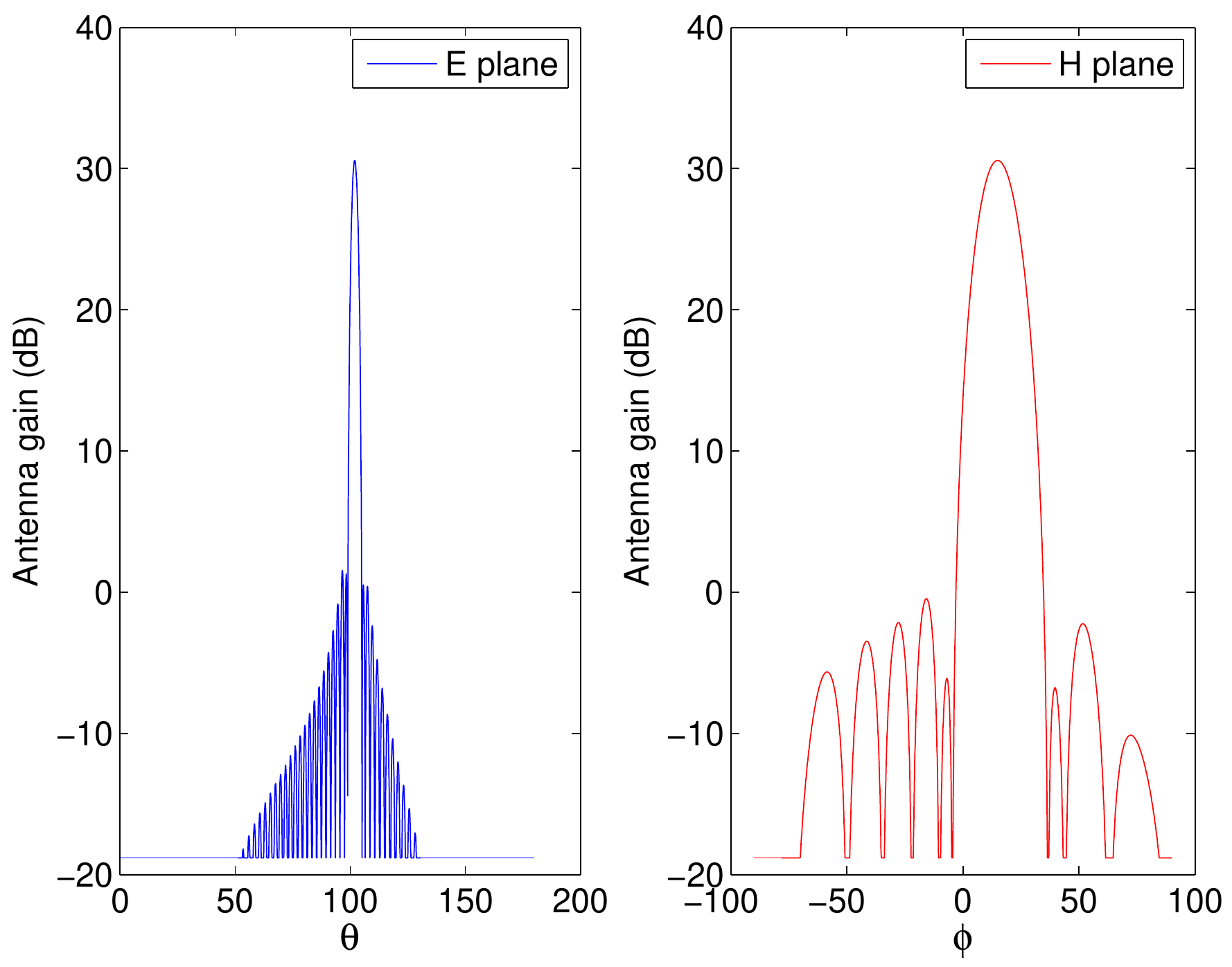}
\caption{\ac{ViS} antenna gain pattern in the E-and H-planes.}
\label{fig:ant_gain}
\end{figure}

	The maximum side-lobe level is given as a constraint (30dB below the maximum gain in the present work). The side lobe level increases with the increase in $\theta_e$  and in $\phi_e$. Hence the antenna design is performed for the maximum planned value of $\theta_e$  and in $\phi_e$. To reach this objective, two levers are available:
	\begin{enumerate}[(i)]
	\item Reducing the distance between the array elements. These should verify the constraint $d_s/\lambda\leq 1; s=x,z$ \label{itm:1}
	\item Increasing the Gaussian tapering, namely the ratio between the extreme and middle amplitudes of the antenna elements in each axis \label{itm:2}
	\end{enumerate}
	where $\lambda$ is the wavelength. Both (\ref{itm:1}) and (\ref{itm:2}) will decrease the side-lobe level and the antenna gain and will increase its main beam-width. Figure \ref{fig:ant_gain} presents the antenna gain in the E- and H-planes for the following parameters: $N_x=10$, and $N_z=40$, $d_x/\lambda=0.5$ and $d_z/\lambda=0.7$. The side lobes' constraints are verified for  $\theta_e \leq 120^\circ$  (namely a tilt up to $30^\circ$) and $|\phi_e| \leq 45^\circ$.

\section{Resource allocation SON Algorithm} \label{sec:son_algo}
	The \ac{SINR} per Hertz of a user $u$ is modeled as follows
	\begin{equation} \label{eq:mac_sinr_novis}
	S_u = \frac{P^s h_u^s}{N_0 + \sum_{c \neq s} P^c h_u^c}
	\end{equation}
	where $P^c$ is the transmit power Per Hertz of \ac{BS} $c$, $h_u^c$ - the signal attenuation from \ac{BS} $c$ to user $u$, $s=\text{argmax}_c P_c h_u^c$ - the best serving cell for user $u$ and $N_0$ the thermal noise per Hertz. The sum over $c \neq s$ accounts for the interference from other \acp{BS}. The frequency diversity is not taken into consideration in the present work.

	The pathloss $h_u^c$ comprises the signal attenuation over the air, the shadowing from the environment and the antenna gains at both the transmitter and the receiver. Fast fading is implicitly taken into account via quality tables which map \ac{SINR} into data rates (averaged over fast fading). The antenna gain at the transmitter is evaluated using Equation \eqref{eq:ant_gain} for a \ac{ViS}. So a better antenna gain will result in a better \ac{SINR}. Let us denote by $m$ and $v$ the indexes related respectively to the macro cell and the \ac{ViS}. The total transmit power available at the macro \ac{BS} $P^0$ is split between the macro cell ($P^m$) and the \ac{ViS} ($P^v$), so $P^0=P^v+P^m$. The \ac{SINR} of a user served by the macro cell in the presence of a \ac{ViS} which reuses the whole bandwidth is
	\begin{equation} \label{eq:mac_sinr_ro}
		S_u = \frac{P^m h_u^m}{N_0 + P^v h_u^v + \sum_{c \neq s} P^c h_u^c}
	\end{equation}
	and the \ac{SINR} of a user served by the \ac{ViS} is
	\begin{equation}\label{eq:vis_sinr_ro}
		S_u = \frac{P^v h_u^v}{N_0 + P^m h_u^m + \sum_{c \neq s} P^c h_u^c}.
	\end{equation}

	In the remainder of the paper especially in the simulation results, we consider only the case where $P^v = P^m = \frac{P^0}{2}$. Equations \eqref{eq:mac_sinr_ro} and \eqref{eq:vis_sinr_ro} clearly show the \ac{SINR} degradation (reduced useful signal, increased interference) when the \ac{ViS} is activated with frequency bandwidth reuse one.

	If instead, the macro cell and the \ac{ViS} operate on disjoint frequencies, then the \ac{SINR} of a macro user is the same as \eqref{eq:mac_sinr_novis} while the \ac{SINR} of a user served by the \ac{ViS} becomes
	\begin{align}
		S_u = \frac{P^v h_u^v}{N_0 + \sum_{c \neq s} P^c h_u^c}
	\end{align}
	where $P^v=P^m=P^0$ since the power available per unit bandwidth does not change. An appropriate choice of the bandwidth sharing proportions is then needed in order to avoid performance degradation. We use the proportional fair sharing criteria which provides a good trade-off between throughput optimization and fairness in resource sharing \cite{bonald_queueinganalysismax_2006},\cite{kushner_convergenceproportionalfair_2004}, \cite{tall_selforganizingstrategies_2014}.

	Denote by $\delta$ the fraction of the frequency bandwidth dedicated to the \ac{ViS} and $\bar{R}_u$ the mean data rate of a user served by either the macro cell or the \ac{ViS} when the other is switched off. The proportional fair utility is defined as
	\begin{equation} \label{eq:ut_func_pf}
	U_{PF}(\delta) = \sum_{u \in \text{ViS}} \log(\delta \bar{R}_u) + \sum_{u \in \text{macro}} \log((1-\delta) \bar{R}_u)
	\end{equation}

	Since the utility function \eqref{eq:ut_func_pf} is concave, maximizing it is a convex optimization problem. Using \ac{K.K.T} conditions for optimality \cite{boyd_convexoptimization_2004}, the optimal value of $\delta$ can be easily derived to be
	\begin{equation} \label{eq:optimal_delta}
		\delta = \frac{N_v}{N_v+N_m}
	\end{equation}
	where $N_v,N_m$ are respectively the number of users in the \ac{ViS} and the macro sector.

	Equation \eqref{eq:optimal_delta} constitutes the self-optimization algorithm used to update the bandwidth sharing proportions between the macro sector and the \ac{ViS} and the update is performed at each event (arrival/departure). It is noted that a general $\alpha$-fair utility \cite{altman_generalizedalphafair_2008} can be used and the optimization problem can be solved using a similar method as in \cite{tall_selfoptimizingstrategies_2015}.

\section{Simulation Results} \label{sec:perf_results}
\subsection{Simulation scenario}
	Consider a trisector \ac{BS} surrounded by 2 rings of interfering macro sites as shown in Figure \ref{fig:net_lay}. In each macro sector, a \ac{ViS} can be activated whenever needed. We consider elastic traffic where users arrive in the network according to a Poisson process, download a file and leave the network as soon as their download is complete. The considered area $A$ is the initial area covered by the central macro \acp{BS}. In order to limit the complexity, slow and fast fading are not taken into account in these simulations and mobility of the users is not explicitly implemented. However the users arrive at random locations in the network.

\begin{figure}[!ht]
\centering
\includegraphics[width=2.8in]{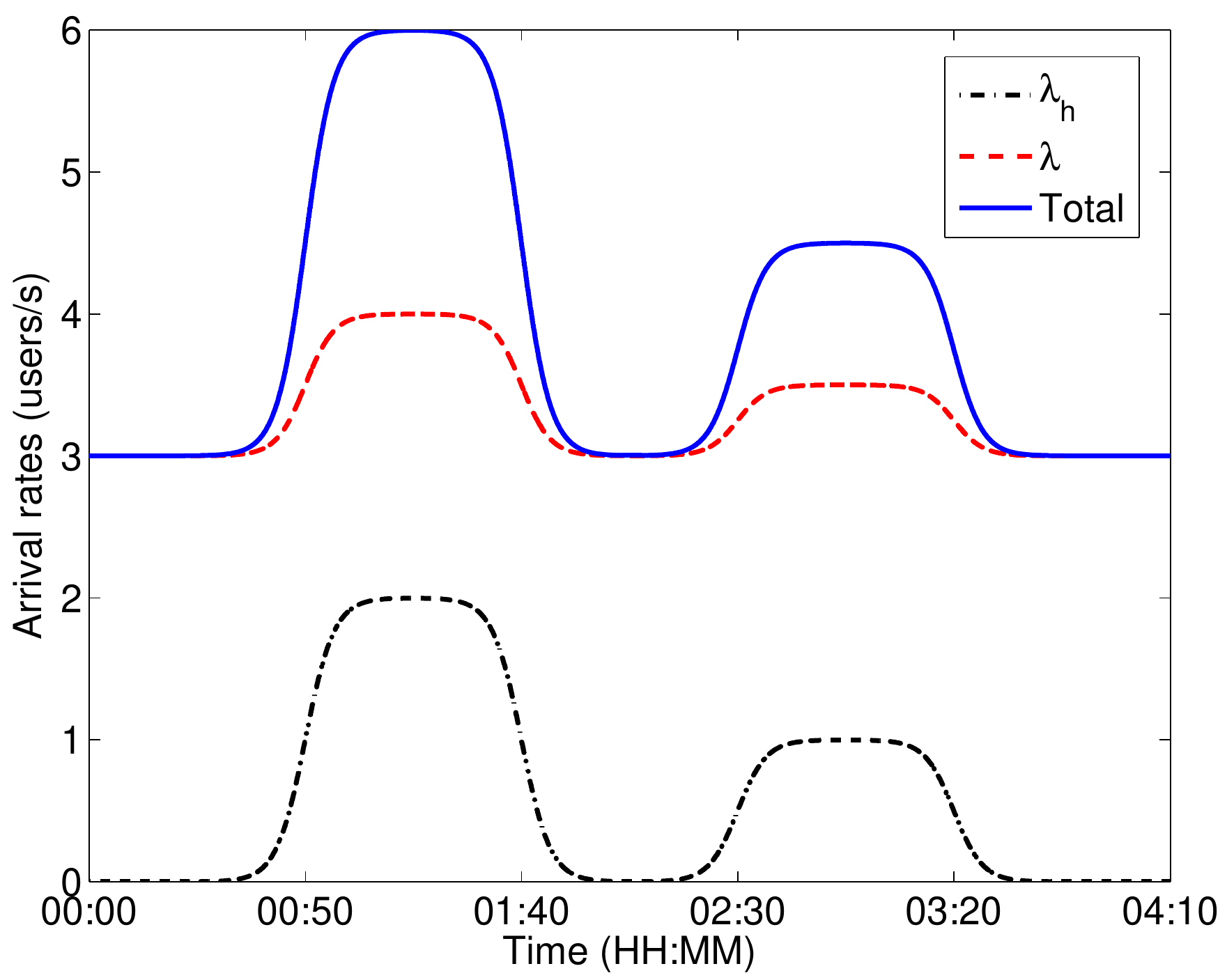}
\caption{Traffic profile over time (HH:MM means hours:minutes)}
\label{fig:traff}
\end{figure}

	Two layers of traffic are superposed: the first one has a uniform arrival rate of $\lambda$ users/s all over $A$, and the second - a uniform arrival rate of $\lambda_h$ users/s in the \acp{ViS} coverage area. These arrival rates evolve over time as shown in Figure \ref{fig:traff} in order to show the effect of the self-optimization algorithm. For example, between 00:50 and 01:40, the hot-spot traffic demand ($\lambda_h$) increases from 0 to 2 users/s. This is close to a realistic scenario where the \acp{ViS}' beams are set to point at the hot-spot areas by adjusting the $\theta_e$, and $\phi_e$ angles.

\begin{table}[!t]
\small
\renewcommand{\arraystretch}{1.3}
\caption{Network and Traffic characteristics}
\label{tab:params}
\centering
\begin{tabular}{|c|c|}
\hline
\multicolumn{2}{|c|}{Network parameters} \\
\hline
Number of macro sectors & 3 \\
\hline
Number of \acp{ViS} & 3 \\
\hline
Number of interfering macros & 2 rings of macro sites \\
\hline
Macro Cell layout & hexagonal trisector \\
\hline
Intersite distance & 500 m \\
\hline
Bandwidth ($B$) & 10MHz \\
\hline
\ac{BS} transmit power & 40W (46dBm) \\
\hline
Scheduler & Round-Robin \\
\hline
Link adaptation model & $B\min(4.4,\log_2(1+\text{SNR}))$ \cite{3gpp_evolveduniversalterrestrial_2012} \\
\hline
\multicolumn{2}{|c|}{Channel characteristics} \\
\hline
Thermal noise & -174 dBm/Hz \\
\hline
Path loss ($d$ in km) & 128.1 + 37.6 $\log_{10}(d)$ dB \\
\hline
\multicolumn{2}{|c|}{Traffic characteristics} \\
\hline
Traffic spatial distribution & uniform + hot-spots \\
\hline
Service type & FTP \\
\hline
Average file size & 3 Mbits \\
\hline
\end{tabular}
\end{table}

\begin{figure}[!ht]
\centering
\includegraphics[width=2.8in]{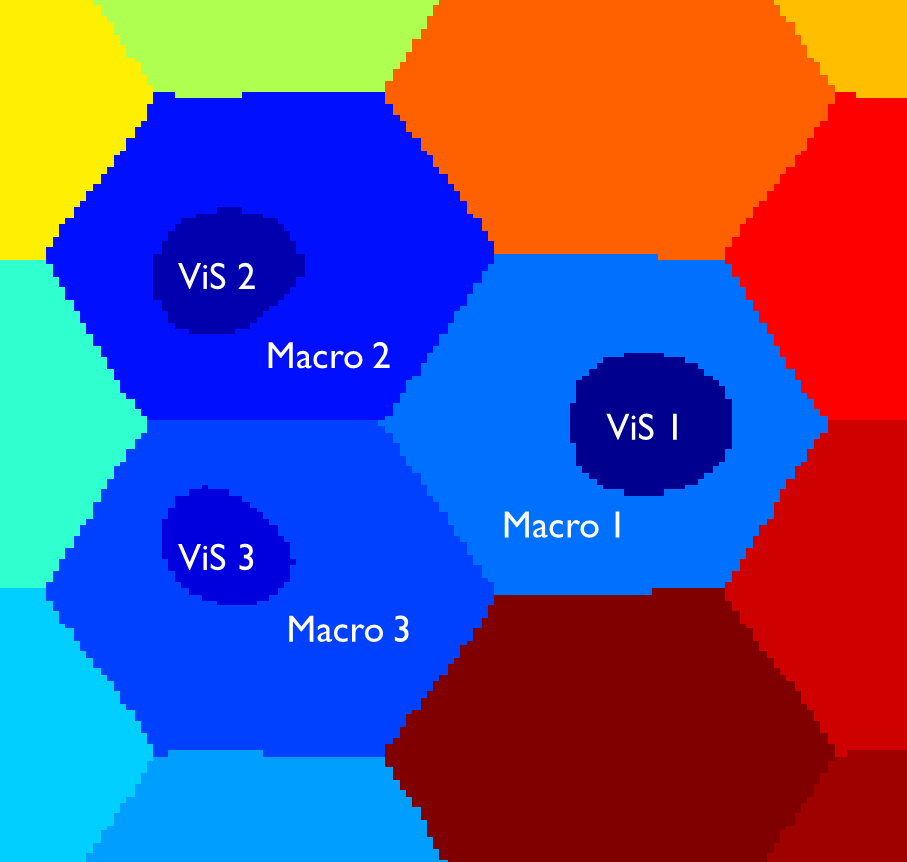}
\caption{Serving cell map}
\label{fig:serving_map}
\end{figure}

	We use the propagation models for all \acp{BS} following \cite[Page 61]{3gpp_evolveduniversalterrestrial_2010} and presented in Table \ref{tab:params} which also summarizes all the simulation parameters. The serving cell map obtained from these parameters is presented in Figure \ref{fig:serving_map}. The parameters used for each \ac{ViS} are summarized in Table \ref{tab:antennas}. The vertical tilt is defined with respect to the horizon and the horizontal tilt has the azimuth of the containing macro sector as reference (see Figure \ref{fig:ant_array}).

\begin{table}[!t]
\small
\renewcommand{\arraystretch}{1.3}
\caption{\ac{ViS}s antenna configurations}
\label{tab:antennas}
\centering
\begin{tabular}{|c|c|c|c|}
    \cline{2-4}
    \multicolumn{1}{c|}{} & VS 1 & VS 2 & VS 3 \\ \hline
    Vertical tilt & 10°  & 11° & 12°  \\ \hline
    Horizontal tilt & 0° & 10° & -15°  \\ \hline
    $N_x$ & \multicolumn{3}{|c|}{10}  \\ \hline
    $N_z$ & \multicolumn{3}{|c|}{40} \\ \hline
\end{tabular}
\end{table}

\subsection{Performance Evaluation}

\begin{figure}[!ht]
\centering
\includegraphics[width=3.5in]{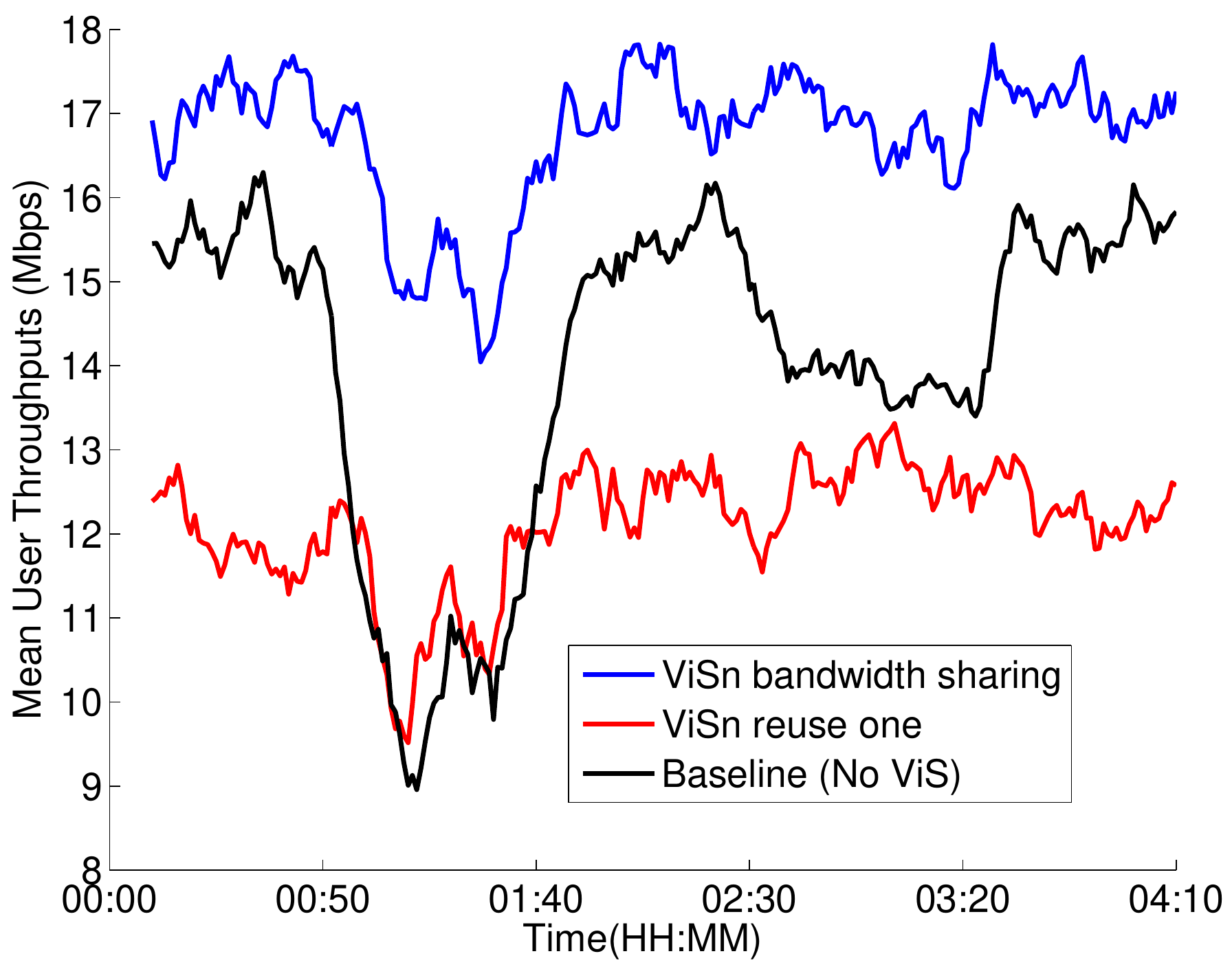}
\caption{Mean user throughputs evolution' over time}
\label{fig:mut}
\end{figure}

\begin{figure}[!ht]
\centering
\includegraphics[width=3.5in]{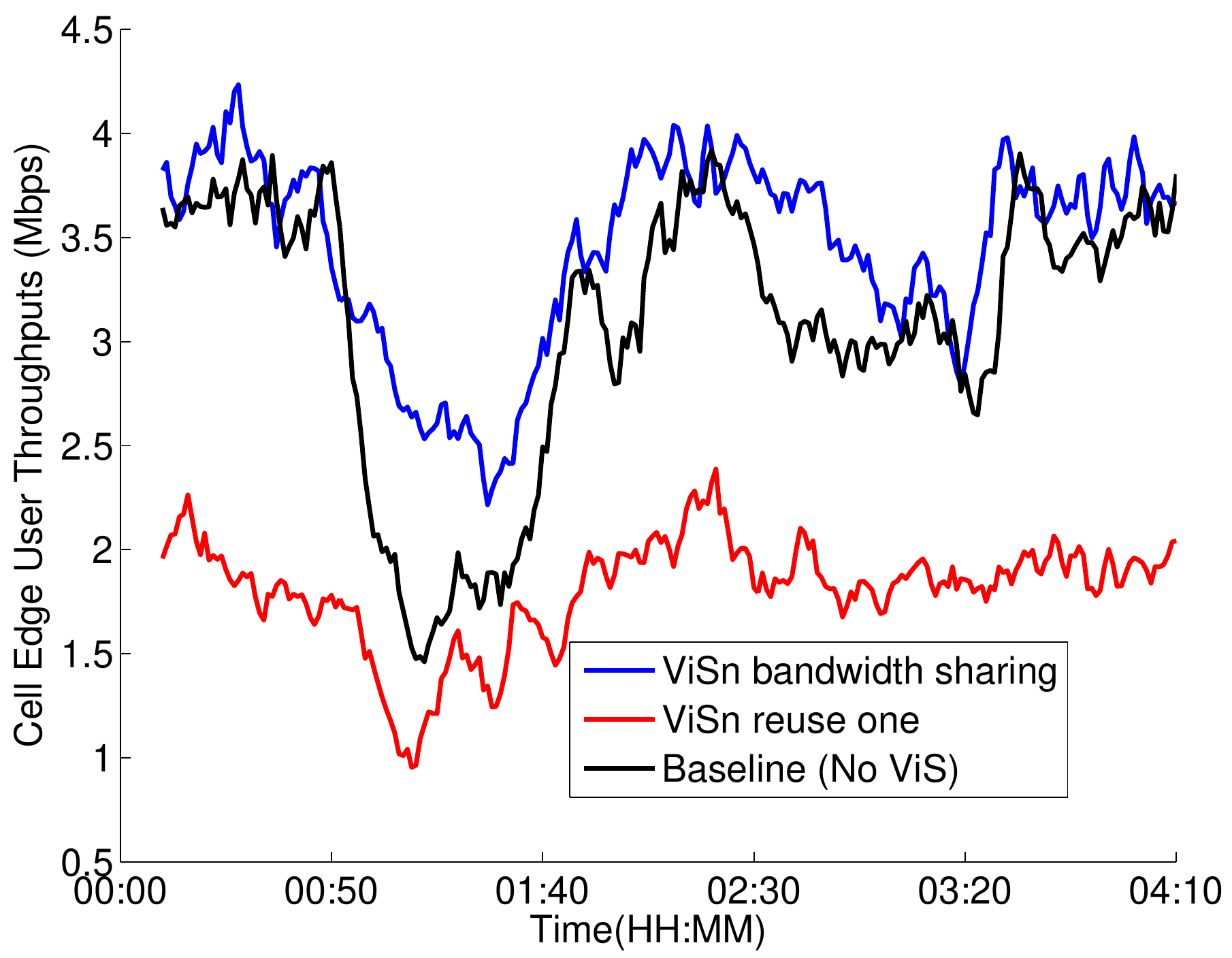}
\caption{Cell edge user throughputs' evolution over time}
\label{fig:cet}
\end{figure}

\begin{figure}[!ht]
\centering
\includegraphics[width=3.5in]{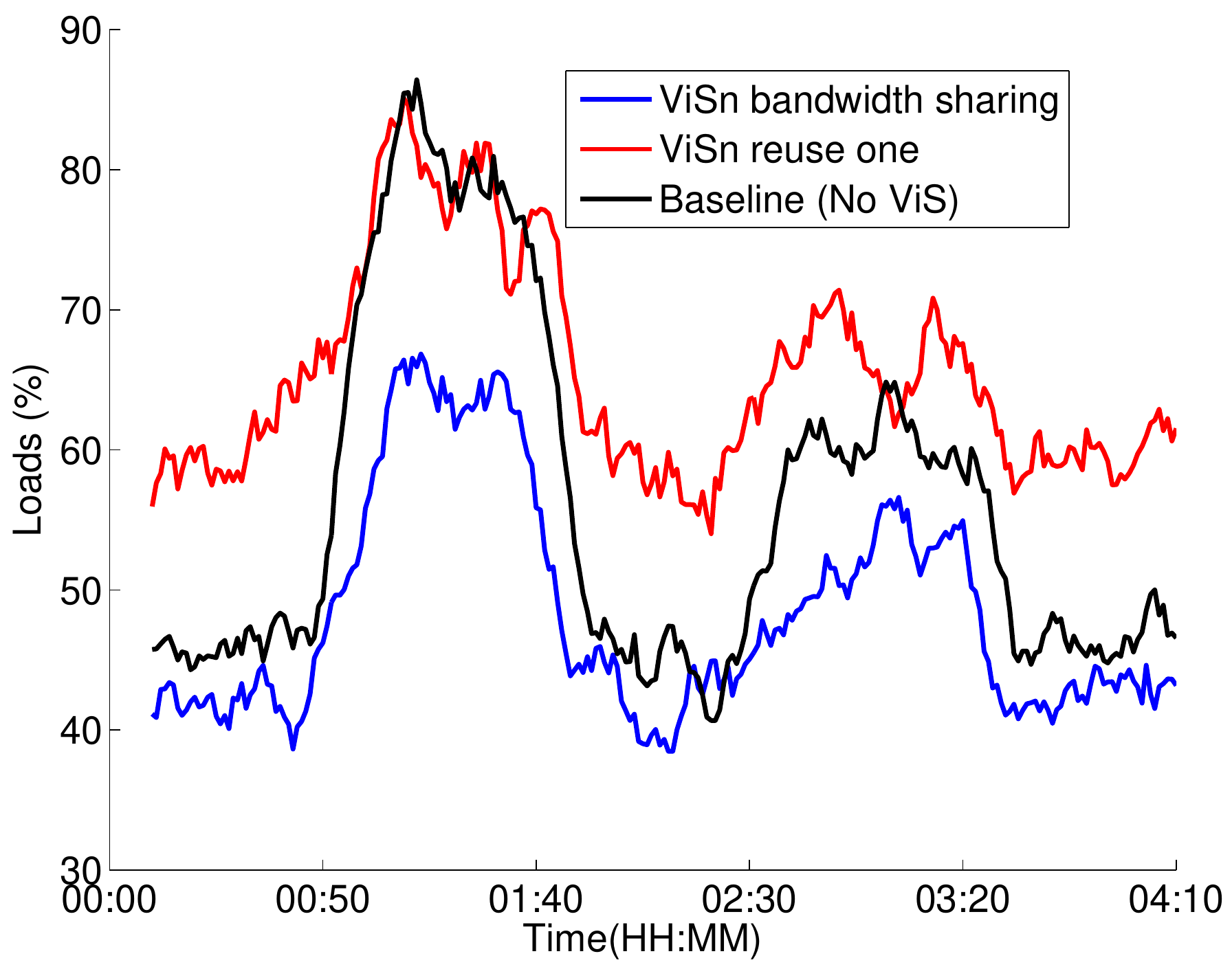}
\caption{Maximum loads' (of all cells, virtual and macro) evolution over time}
\label{fig:rho}
\end{figure}

\begin{figure}[!ht]
\centering
\includegraphics[width=3.5in]{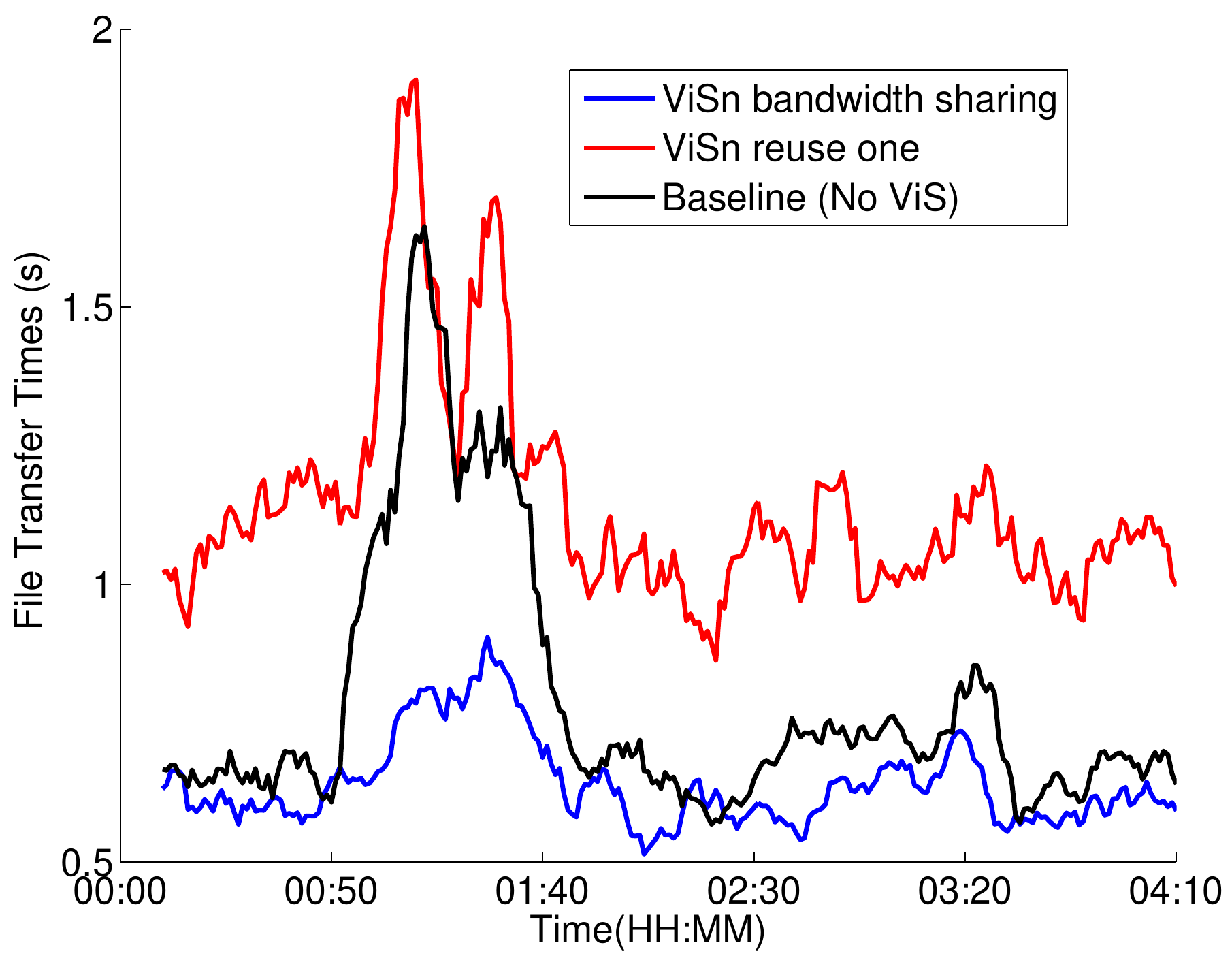}
\caption{File transfer time' evolution over time}
\label{fig:ftt}
\end{figure}

	We evaluate the \ac{MUT} (Figure \ref{fig:mut}), the \ac{CET} (Figure \ref{fig:cet}), the maximum loads (Figure \ref{fig:rho}) and the \ac{FTT} (Figure \ref{fig:ftt})  for three different cases:
	\begin{itemize}
	\item Baseline (black in Figures): this is the reference case in which no \ac{ViS} is present, so the macro sectors serve all the traffic as they would traditionally.
	\item \ac{ViSn} reuse one (red in Figures): in this case, the \acp{ViS} are deployed with a full reuse of the bandwidth. The macro and virtual sectors share equally the available transmit power.
	\item \ac{ViSn} bandwidth sharing (blue in Figures): the \acp{ViS} are also enabled in this case but the total bandwidth is shared between the macro cell and the \ac{ViS} in its coverage area. The bandwidth sharing proportions are dynamically optimized using \eqref{eq:optimal_delta}.
	\end{itemize}
	It is noted that the \ac{CET} refers to the 5th percentile throughput, so it will correspond generally to users at the macro cell edge in our scenario (no interference between macro cell and \ac{ViS}).

	The numerical results show that deploying the \ac{ViS} with full reuse of the bandwidth degrades performance over the baseline (No \ac{ViS}) except for sufficiently high loads. Indeed the \ac{CET} and the \ac{FTT} of the baseline is always the same or better than those of the reuse one case. It is only between 00:50 and 01:40 that the \ac{MUT} of the baseline is slightly worse than that of reuse one (see Figure \ref{fig:mut}), and it can be seen in Figure \ref{fig:rho} that the mean load at this time is over 75\% for both baseline and reuse one.

	The reuse one case degrades performance because of its worse \ac{SINR} (reduced power due to its split between macro and virtual cells, and macro-virtual cells mutual interference). The \ac{CET} of reuse one is still worse  This scheme is then only useful at very high loads (over 85\%). It is noted that similar results have been obtained in \cite{tall_selfoptimizingstrategies_2015} for \ac{VeSn}.

	Deploying the \ac{ViS} with bandwidth sharing is shown to provide the best performance during the whole simulation period for different load conditions as shown by all performance indicators in Figures \ref{fig:mut}, \ref{fig:cet} and \ref{fig:ftt}. Even the loads (Figure \ref{fig:rho}) are lower suggesting that deploying \ac{ViS} with bandwidth sharing provides a higher capacity.

	The higher gain of the \ac{ViS} antenna improves its \ac{SINR} over the baseline case. Moreover, the bandwidth sharing enables the two cells (macro sector and \ac{ViS}) to serve their traffic without mutual interference and a better \ac{SINR} compared to a \ac{ViS} deployed with full bandwidth reuse. It is noted that the bandwidth reuse one is expected to provide better performance when the loads approach 100\%.

\section{Conclusion} \label{sec:conclusion}
	This paper has developed a model for virtual sectorization, encompassing the antenna design and the \ac{SON} algorithm for frequency bandwidth allocation. Using an antenna array, a focused beam can be created to cover a small area delimiting for example a traffic hot-spot. The focused beam provides a higher antenna gain thus a higher \ac{SINR}, but its performance can be limited by the macro-cell interference. A simple proportional fair based \ac{SON} algorithm is used to share the total bandwidth between the macro cell and the \ac{ViS} in its coverage area, thus eliminating the mutual interference between them. The numerical results demonstrate the significant performance gain brought about by self-optimized \ac{ViSn}.

	\ac{ViSn} has a clear advantage with respect to \ac{VeSn}, since it allows to generate a sector anywhere in the macro-cell coverage zone. When the coverage area of the \ac{ViS} is of the order of 20 percent of the macro-cell area, the \ac{ViS} antenna can have a reasonable size, of the order of $2.4m \times 1m$ for 2.6 GHz (i.e. typical \ac{LTE} frequency) and can be viewed as a 4G technology. If one aims at achieving higher antenna gain covering smaller cell size, the number of radiating elements of the antenna array will considerably increase, and therefore higher operating frequencies are required. In this case, the \ac{ViSn} should be considered rather as 5G technology.

\section*{Acknowledgment}
	The research leading to these results has been partially carried out within the FP7 SEMAFOUR project and has received funding from the European Union Seventh Framework Programme (FP7/2007-2013) under grant agreement no 316384.

\bibliographystyle{IEEEtran}
\bibliography{main}

\end{document}